\documentclass[prb,twocolumn]{revtex4}
\usepackage{epsfig}
\usepackage{graphics}
\renewcommand{\vr}{\vec{r}}
\newcommand{\ts}{\tilde{\sigma}}
\newcommand{\psfigur}[2]{\centerline{\epsfig{file=#1.eps,width=#2cm}}}

\newcommand{\sdu}{\langle n_{i\uparrow}\rangle}
\newcommand{\sdd}{\langle n_{i\downarrow}\rangle}

\begin{document}

\title{Charge transport through image charged stabilized states in a
single molecule single electron transistor device}
\author{Per Hedeg\aa rd}
\affiliation{Niels Bohr Institute, University of Copenhagen,
Universitetsparken 5, DK-2100 Copenhagen, Denmark}
\author{Thomas Bj\o rnholm}
\affiliation{Nano-Science Center and Department of Chemistry,
University of Copenhagen, Universitetsparken 5, DK-2100 Copenhagen,
Denmark}
\begin{abstract}
The present paper gives an elaborate theoretical description of a
new molecular charge transport mechanism applying to a single
molecule trapped between two macroscopic electrodes in a solid state
device. It is shown by a Hubbard type model of the electronic and
electrostatic interactions, that the close proximity of metal
electrodes may allow electrons to tunnel from the electrode directly
into a very localized image charge stabilized states on the
molecule. Due to this mechanism, an exceptionally large number of
redox states may be visited within an energy scale which would
normally not allow the molecular HOMO-LUMO gap to be transversed.
With a reasonable set of parameters, a good fit to recent
experimental values may be obtained. The theoretical model is
furthermore used to search for the physical boundaries of this
effect, and it is found that a rather narrow geometrical space is
available for the new mechanism to be effective: In the specific
case of oligophenylenevinylene molecules recently explored in such
devices several atoms in the terminal benzene rings need to be at
van der Waal's distance to the electrode in order for the mechanism
to be effective.  The model predicts, that chemisorption of the
terminal benzene rings too gold electrodes will impede the image
charge effect very significantly because the molecule is pushed away
from the electrode by the covalent thiol-gold bond.
\end{abstract}
\maketitle

\section{Introduction}
Charge transfer processes of single electrons over \AA ngstr\"om
distances between well defined molecular donor and acceptor moieties
are ubiquitous in nature and they are extensively
studied\cite{no1,no2}. Classical examples include the photosynthetic
reaction center, redox active enzymes and Kreutz-Taube complexes
\cite{no1,no2}. For all of these processes, the contact to a
temperature bath, molecular vibrations (phonons) and the
reorganization of the molecular medium surrounding the CT process
all play significant roles as expressed by Marcus theory\cite{no3},
\begin{equation}
k = A \exp(- \Delta G^*/k_BT)
\end{equation}
where $k$ is the rate of charge transfer and $A$ denotes the
preexponential rate factor. The free energy,  $\Delta G^* = (
\lambda/4)(1 + \Delta G_0/\lambda )^2$, where $G_0$ denotes the
thermodynamic driving force of the reaction and $\lambda$ the
reorganization energy introduced by Marcus. In typical cases the
reorganization energy will involve ions, and dipoles of the
surrounding solvent as well as internal reorganization of the
molecular CT-system.

By exchanging the donor or acceptor moiety with an electrode,
centuries of electrochemical studies have advanced the understanding
of charge transfer between a metal electrode and a redox active
species in electrolyte solutions\cite{no4}. Very recent experiments
employing scanning tunneling microscopes under electrochemical
control have further extended these studies to the single molecule
level\cite{no5,no6}.

As an extension of this historical development this paper also
describes electron transfer processes on a single molecule level,
but in the absence of a solvent. The experiment is based on a solid
state devices in which a single molecule has been trapped between
two metal electrodes in ultra high vacuum at 4 Kelvin\cite{no7}.
Compared to previous measurements, temperature effects are hence
suppressed and solvent and electrolyte effects are naturally absent.
The presence of two metal electrodes in constant contact with the
molecule, however, produces significant new effects resulting in a
charge transport mechanism very different from that observed in
classical photoinduced or electrochemically driven charge transfer
systems. As described and analyzed in the following, we claim to
have discovered a new charge transport channel mediated by image
charge stabilized states lying in the band gap between the
corresponding redox states in solution. The present paper emphasizes
and expands on the theoretical treatment of this new phenomenon
which was introduced in a recent paper describing the experimental
possibility to visit several single molecule redox states in such a
solid state device\cite{no7}. Since two historically rather
segregated fields of science, ``mesoscopic physics'' and ``redox
chemistry'', are now coming together at the single molecule scale,
we have chosen to describe the experiment reported in Ref. 7 in some
detail followed by a introduction to the typical "mesoscopic
physics" description of the charge transport phenomena. Following
this, we describe the detailed theory for the image charged states
and finally make some prediction for how changes in molecular
structure and device geometry should affect future experiments.

\section{Single molecule - single electron transistors} A transistor is an
electronic device with 3 leads, source, drain and gate. The current
from source to drain is controlled by either the voltage or the
current through the gate. In the examples we shall consider in the
present paper it is the voltage on the gate that is used as a
control parameter. The concept is so broad that the electrochemical
cell also is covered. The current runs from working electrode to the
counter electrode and is controlled by the reference electrode. Our
main emphasis is on the solid state device sketched in figure 1.
Here an island of either a small metallic particle or a single
molecule
--- hence the name ``Single molecule transistor'' --- is placed between
the source and and drain electrodes. These electrodes are typically made from gold.
The construction is grown on top of the third gate electrode with an insulating
layer in between to prevent any current to move through the gate. The island
and the leads are hence only capacitively coupled to the gate.

\begin{figure}
\centerline{\includegraphics*[width=70mm]{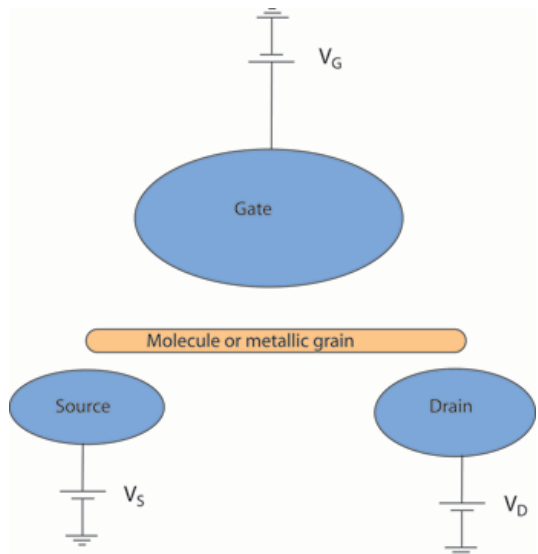}}
\caption{Schematic Single Electron Transistor. The three leads,
source, drain and gate are kept at constant potentials. The central
island (metallic grain or molecule) has a well defined charge.}
\end{figure}

The experiments we are going to consider here are carried out at 4
Kelvin, so quantum effects need to be taken into account. The most
basic process is the transfer of an electron from one of the leads
to the central island. At very low temperatures this is a quantum
tunneling process, which is characterized by a tunneling matrix
element. This corresponds to an energy, $\Gamma$, or via the
Heisenberg uncertainty principle to time scale $\tau =
\hbar/\Gamma$. The property of the transistor depends on $\Gamma$
compared to other energies of the problem. The most relevant other
energy scale is the change in electrostatic energy, $\Delta E_c$, as
the electron is moved to the island. In the so-called ``strong
coupling'' limit, where $\Gamma \gg \Delta E_c$, the electron will
be delocalized and is described by a wavefunction, which has weights
in both leads and on the island. The total charge on the island is
simply not a good quantum number. In this limit standard
Hartree-Fock or LDA calculations provide a good starting point for
quantitative determination of e.g.~the conductivities of the
transistor. The other extreme, so-called ``weak coupling'', is
characterized by $\Gamma$ being the smallest relevant energy
parameter of the problem. The important states of the system has the
total charge of the island as a good quantum number. Electron
transport is a very weak process, which can be described in
perturbation theory, with the above tunneling processes as the
perturbation. This is also the typical regime encountered for
molecular charge transfer processes.

An important notion is that of ``quantum coherence''. In principle
all degrees of freedom, electrons, phonons, molecule position, etc.,
should be included when solving Schr\"odinger's equation. In
practice only a subsystem, e.g.~ the electrons, are considered. This
will often, and certainly at low temperatures, be a good
approximation, but after some characteristic time, $\tau_{coh}$ the
approximation breaks down and the effect of the environment
(phonons, etc.) cannot be neglected. One way of describing this is
to say, that the environment make a ``measurement'' on the electron
system, and ``collapses'' it into one of a set of states with a
certain probability. The Schr\"odinger equation for the electrons
now takes over, with the collapsed state as its new initial state.
The states to which the electrons collapses are particularly robust,
and are the states which are relevant in the classical limit, where
intricate quantum interference effects are suppressed. The nature of
the coupling to the environment is such that these ``classical''
states has the total charge on the island as a good quantum number.
The coherence time $\tau_{coh}$ is strongly temperature dependent,
rapidly decreasing with increasing temperature, and at not to high
temperatures a master equation model, where one is only considering
classical states and rates of transfer between these, is very
effective at giving a good quantitative description.

The quantum states relevant for weak coupling are in fact
``classical'' (with a definite number of electrons on the central
island), and the above mentioned master equation description is
adequate. The electrons are being transported through the transistor
through a series of ''classical'' states one at a time, hence the
name ``Single Electron Transistor'' (SET). Much of current research
on Single Molecule, Single Electron Transistors is concerned with
bridging the gap between the two limits of strong and weak coupling.
A simple rule of thumb says, that if the conductivity of the device
is much smaller than the conductance quantum $e^2/\hbar$ then the
transistor is in the weak coupling limit, while for a device with a
much larger conductivity is in the strong coupling limit.

The experiments we are focussing on in this paper is certainly in
the weak coupling limit. We will hence be emphasizing the nature of
the ''classical states'', which are not as simple as one might at
first think.

In each of the steps involved in the transport of an electron from
source to drain energy conservation need to be satisfied. In the
case where the central island is in fact a metal droplet, the
determination of the total energy of a configuration is really an
exercise in classical electrostatics, which we shall briefly review
below. When the island becomes very small, like e.g.~a single
molecule, the determination of the total energy is a full quantum
mechanical problem, where only part of the problem is
electrostatics. What is needed in this case is a combination of
classical electrostatics and quantum chemistry, and this will be the
main result of this paper.

\subsection{Orthodox SET theory}
In the orthodox theory of Single Electron Transistors (see
e.g.~\cite{no11}) all electrodes are considered to be classical
conductors which, when charged, are coupled via a number of
capacitances. Typically there is a capacitance, $C_s$, coupling
island and the source electrode, and capacitances , $C_d$ and $C_g$,
coupling the island the the drain and gate electrodes respectively.
Implicity in such a model is the notion that a charge, $Q$ on the
island is split in three, $Q_s$, $Q_d$ and $Q_g$, which are coupling
only to their respective countercharges $-Q_s$, $-Q_d$ and $-Q_g$ on
the other electrodes. In reality all charges are of course coupled,
and the more so in the relevant geometry of a very small island,
where e.g.~$Q_s$ on the island and $-Q_s$ on the source electrode
are physically very close, and both will couple to e.g.~$-Q_g$ on
the gate electrode. We hence set up a general classical scheme,
where all couplings are treated on equal footing.

Assume now, that we have a set of conducting electrodes with charges
$Q_i$. Each electrode is an equipotential with with electrical
potential, $V_i$. Poisson's equation ensures, that there is a linear
relation between charges and potentials:
\begin{equation}
\label{VvsQ} V_i = \sum_j P_{ij} Q_j,
\end{equation}
where the coefficients, $P_{ij}$ are entirely determined by the
geometry of the system. The total electrostatic energy of the system
is in this case given by
\[U = \frac{1}{2}\sum_{ij}P_{ij}Q_i Q_j\]
In the actual physical system the charges on the source, drain and
gate electrodes are not given, rather it is the electrical
potentials on these that are given and maintained e.g.~by attached
batteries. The charge on the island is still considered fixed, and
the potential of the island is a variable depending on the
potentials of the other electrodes and the charge on the island. Let
us denote this charge, $Q_0$, and use Greek letters to denote the
source, drain and gate electrodes. Regarding the potentials
$V_\alpha$ and the charge $Q_0$ as independent variables we can
solve (\ref{VvsQ}) with respect to the electrode charges and the
island potential. We get for the electrode charges
\begin{equation}\label{Qalpha}
Q_{\alpha} = \sum_\beta C_{\alpha\beta} (V_\beta-P_{\beta 0} Q_0),
\end{equation}
and for the island potential
\begin{equation}
V_0 = \sum_\alpha f_\alpha V_\alpha + \frac{Q_0}{C}.
\end{equation}
Here the capacitance matrix $C_{\alpha\beta}$ is the inverse of
$P_{\alpha\beta}$. The dimensionless parameters $f_\alpha$, and the
island capacitance $C$ are given by
\begin{eqnarray}\label{fandC}
    f_\alpha &=& \sum_{\beta} P_{0\beta}C_{\beta\alpha}\nonumber\\
    C &=&
    (P_{00}-\sum_{\alpha\beta}P_{0\alpha}C_{\alpha\beta}P_{\beta
    0})^{-1}.
\end{eqnarray}
The total electrostatic energy is in terms of these variables
\begin{equation}\label{Utot}
    U = \frac{1}{2}C_{\alpha\beta} V_\alpha V_\beta +
    \frac{Q_0^2}{2C}.
\end{equation}
Assume now, that the charge on the island is changed an amount
$\delta Q_0$. Then according to (\ref{Qalpha}), the charge on the
leads will change an amount
\begin{equation}
\delta Q_\alpha = - \sum_\beta C_{\alpha\beta}P_{\beta 0} \delta Q_0
= -f_\alpha \delta Q_0.
\end{equation}
This gives a very simple interpretation of the three $f_\alpha$
parameters, as the fraction of counter charge on the electrodes
induced by the extra charge on the island.

For a current to flow from source to drain there are two possible
scenarios: a) first an electron moves from source to island, and
then from island to drain or b) first an electron moves from island
to drain and then an electron moves from source to island. In
general the battery work involved in the first step of scenario a)
will be
\begin{eqnarray}
W_{bat}^{(a)} &=& eV_S(1-f_S) - eV_D f_D - eV_G f_G \nonumber \\
&=& e V (1-f_S+f_D)/2 - e (V_G f_G - V_{SD}(1-f_S-f_D)),
\end{eqnarray}
where we have introduced the average of the source and drain
potentials, $V_{SD}$, and their difference, $V = V_S-V_D$. Likewise
the work provided in the first step of b) where an electron moves
from island to drain is
\begin{equation}
W_{bat}^{(b)}= e V (1+f_S-f_D)/2 + e (V_G f_G - V_{SD}(1-f_S-f_D)).
\end{equation}
The total battery work in both scenarios is the sum of
$W_{bat}^{(a)}$ and $W_{bat}^{(b)}$, which is $eV$, so if
$eV_S>eV_D$ then the overall process is energetically possible. The
first step, however, is not necessarily possible.

Consider a situation where are $n$ electrons on the island,
i.e.~$Q_0=ne$, and where the average potential of the source and
drain potentials is equal to the ground potential, i.~e. zero. The
change in electrostatic energy in process a) will be
\begin{equation}
\Delta U_a = \frac{e^2}{2C}((n+1)^2-n^2) = \frac{e^2}{2C}(2n+1).
\end{equation}
The battery work in the first step will be
\begin{equation}
W^{(a)}_{bat} =e V (1+f_S-f_D)/2- f_G eV_G.
\end{equation}
If $W^{(a)}_{bat}\ge \Delta U_a$ i.e.~if
\begin{equation}
eV > 2\frac{\Delta U_a + f_G eV_G}{1+f_S-f_D}
\end{equation}
then the process is possible, and current will flow. If $eV_G < -
\Delta U_a/f_G = - \frac{e^2}{2C f_G}(2n+1)$ then the step will
happen even for $eV=0$. This means that an extra electron will move
permanently to the island, and the analysis should be repeated with
$Q_0=(n+1) e$.

In the first step in the b) process electrostatic energy will be
\begin{equation}
\Delta U_b = \frac{e^2}{2C}((n-1)^2-n^2) = -\frac{e^2}{2C}(2n-1).
\end{equation}
The battery work will now be
\begin{equation}
W^{(b)}_{bat}=e V (1+f_S-f_D)/2 + eV_G f_G.
\end{equation}
If $W^{(b)}_{bat}>\Delta U_b$, i.e.~if
\begin{equation}
eV > 2\frac{\Delta U_b - f_G eV_G}{1-f_S+f_D}
\end{equation}
we will have current flowing according to scenario b). For $eV_G >
\Delta U_b/f_G = -\frac{e^2}{2C f_G}(2n-1)$ then the first step of
moving an electron from island to drain will be permanent, and we
should replace $Q_0$ by $(n-1)e$.

The situation is summarized in the following figure.

\begin{figure}
\psfigur{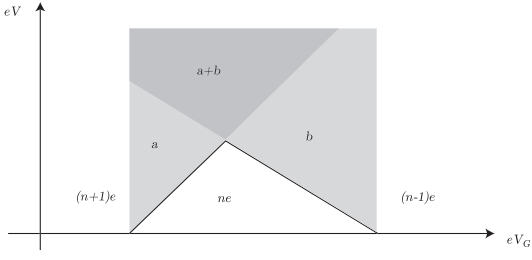}{7} \caption{Coulomb blockade diamond. The single
electron transistor is only transporting current in the shaded
areas. The horizontal axis is the gate potential; the vertical axis
is the source-drain potential difference.}
\end{figure}

The width of the white diamond is given by $\Delta eV_G =
\frac{e^2}{f_G C}$, while the two slopes are given by $\alpha_1 =
\frac{2 f_G}{1+f_S-f_D}$ and $\alpha_2 = - \frac{2f_G}{1-f_S+f_D}$.
From these parameters, which can be determined experimentally, we
can read of the following electrostatic parameters:
\begin{eqnarray}\label{fG}
f_G = \frac{\alpha_1\alpha_2}{\alpha_1-\alpha_2} \nonumber \\
f_D - f_S = \frac{\alpha_1+\alpha_2}{\alpha_1-\alpha_2}
\nonumber\\
C = \frac{e^2}{f_G\Delta eV_G}.
\end{eqnarray}
It is easy to show, that $f_G$ is the ratio of the height of the
diamond to its width, which simplifies the determination of this
important parameter if there is experimental access to the full
diamond.

\subsection{Beyond orthodox SET. Single molecule transistors} The
orthodox theory of Single Electron Transistors is based on a
classical electrostatic conductor model of both electrodes and the
central island. In a Single Molecule Transistors, the central island
is replaced by a molecule, who's electronic structure and total
energy is determined by quantum mechanics. This e.g.~means that the
excess energy in the gray areas of Fig.~2, cannot necessarily be
deposited in the molecule due to the discrete nature of the
molecular energy levels. Another problem, which is particularly
acute in the recent experiments on Single Molecule Transistors where
the molecule is OPV5. Here the coupling between electrons on the
molecule and induced charges in the electrodes --- the so-called
image charges --- will significantly change the electronic structure
of the molecule, and hence affect the transport through the
transistor. These effects call for a generalization of the orthodox
model, and this is the purpose of the following discussion.

The charge density, $\rho_M(\vr)$, on the molecule will induce
charges in the electrodes and change potential of these. In
principle $\rho_M(\vr)$ is a quantum mechanical operator, but in
what follows we shall use the Hartree approximation for this, since
the effects we are after can indeed be captured in this
approximation. The metallic electrodes are still described as
classical electrostatic conductors. For good conductors with a high
electron density this is certainly a good approximation, as
discussed by Lang and Kohn\cite{no9}.

The potential of electrode with label $\alpha$ ($\alpha$ still
referring to either source, drain or gate), now becomes
\begin{equation}\label{newPotential}
    V_\alpha = \sum_\beta P_{\alpha\beta}Q_\beta + V^M_\alpha,
\end{equation}
where the parameters $P_{\alpha\beta}$ are characteristic for the
geometry of the electrodes {\em without} the molecule. $V^M_\alpha$
represent the coupling between the molecular charge density and the
so-called {\em canonical} charge density, $\ts_\alpha(\vr)$, of a
unit charge added to electrode $\alpha$:
\begin{equation}\label{vm}
    V^M_\alpha
    =\frac{1}{4\pi\epsilon_0}\int\frac{\ts_\alpha(\vr)\rho_M(\vr')}{|\vr-\vr'|}\,da\,dv'.
\end{equation}
The total electrostatic --- or Coulomb energy --- in this case now
becomes
\begin{eqnarray}\label{CoulombEnergi}
    U &=& \frac{1}{2}\sum_{\alpha\beta} P_{\alpha\beta}Q_\alpha
    Q_\beta + \sum_\alpha Q_\alpha V^M_\alpha \nonumber \\
    && + \frac{1}{8\pi\epsilon_0}\int
    \frac{\rho_M(\vr)\rho_M(\vr')}{|\vr-\vr'|}\,dv\,dv' \nonumber \\
    && + \frac{1}{8\pi\epsilon_0}\int
    \frac{\rho_M(\vr)\sigma_M(\vr')}{|\vr-\vr'|}\,dv\,da',
\end{eqnarray}
where $\sigma_M(\vr)$ is the charge density induced on the surfaces
of the electrodes. In contrast to the canonical charge densities,
this ``image'' charge density is quite localized to a region on the
electrodes close to the molecule. In fact, $\sigma_M(\vr)$ is
linearly related to $\rho_M(\vr)$, so we can write
\begin{equation}\label{imagecharge}
    \sigma_M(\vr) = - \int K(\vr,\vr') \rho_M(\vr') dv',
\end{equation}
where the kernel $K(\vr,\vr')$ is given by the geometry of the
system. We use a minus in the definition, since $\sigma_M(\vr)$ is
describing an image charge. If we insert (\ref{imagecharge}) into
(\ref{CoulombEnergi}) we get the final version of the total
electrostatic energy
\begin{eqnarray}\label{finalCoulombEnergi}
U &=&\frac{1}{2}\sum_{\alpha\beta} Q_\alpha Q_\beta P_{\alpha\beta}
+ \sum_\alpha Q_\alpha V_\alpha^M
\nonumber \\
&& + \frac{1}{8\pi\epsilon_0} \int
\rho_M(\vr)F(\vr,\vr')\rho_M(\vr') dv dv',
\end{eqnarray}
where the kernel $F(\vr,\vr')$ is given by
\begin{equation}
F(\vr,\vr') = \frac{1}{|\vr-\vr'|} - \int
\frac{K(\vr'',\vr')}{|\vr-\vr''|} dv''.
\end{equation}

So far we have considered the charge distributions as given. They
are determined by several factors. The total charge on the molecule
can be considered fixed and given, but the actual distribution is
determined in a fully quantum mechanical calculation. The charge
distribution on the molecule is given in terms of the one-electron
wavefunctions (e.g.~the wavefunctions appearing in the Kohn-Sham
equations of the Density Functional Theory):
\begin{equation}
\rho_M(\vr) = e \sum_{n} \psi^*_n(\vr) \psi_n(\vr)+ \rho_N(\vr),
\end{equation}
where $e$ is the electron charge and $\rho_N(\vr)$ is the charge
density of the nuclei. The potential energy term in Schr\" odinger's
equation includes the potential from the other charges in the
problem and can be obtained as
\begin{eqnarray}\label{SchrPotential}
V_C (\vr)\psi_n(\vr) &=& \frac{\delta U}{\delta \psi^*_n(\vr)}
\nonumber
\\
&=& \sum_\alpha \frac{Q_\alpha e}{4\pi\epsilon_0}\int
\frac{\ts_\alpha(\vr')}{|\vr'-\vr|}da' \psi_n(\vr)
\nonumber \\
&& + \frac{e}{4\pi\epsilon_0}\int \rho_M(\vr') F(\vr',\vr)
\psi_n(\vr),
\end{eqnarray}
where the first term is the interaction with the charges on the
electrodes, and the second is the Hartree term of the molecule plus
interaction with image charges. In deriving this result we have
assumed that the electron system of the metal electrodes are very
fast degrees of freedom, which instantaneously respond to changes in
the electron distribution of the molecule.

The total energy of the molecule, i.e. the eigenvalue of the many
body Schr\" odinger equation will hence {\it include} the two last
terms of the total electrostatic energy from equation
(\ref{finalCoulombEnergi}). The total energy of the entire
configuration, i.e.~including both electrostatic energy stored in
the charged electrodes and the total energy of the molecule in this
particular environment becomes
\begin{equation}\label{Etot}
    U_{tot} = \frac{1}{2}\sum_{\alpha\beta} P_{\alpha\beta} Q_\alpha
    Q_\beta + E_{tot}(n),
\end{equation}
where the last term is the total energy of the molecule, {\em
including} coupling to electrodes and image charges.

In the physical situation we have in mind, the electrodes are kept
at fixed potential --- maintained by ``batteries''. This means that
the charge distributions $\sigma_\alpha(\vr)$ are dependent
variables, determined by the geometry, by the fixed potentials
$V_\alpha$, and by the molecular charge, $\rho(\vr)$. As in the
orthodox theory we shall eliminate the total charges $Q_\alpha$ from
the problem. We get
\begin{equation}\label{inversion}
Q_\alpha = \sum_\beta C_{\alpha\beta}(V_\beta-V_\beta^M).
\end{equation}
The potentials $V_\beta^M$ depends on the actual charge distribution
on the molecule, which is not easily accessible. In order to proceed
further we need to make some simplifying assumptions. If the
electrodes are quite far from the molecule, or if the electrodes
have small potential differences, then the electrical potential due
to the electrode charges will be almost constant in the region of
the molecule, and the contribution to the total electrostatic energy
from this potential is well approximated by
\begin{equation}\label{approx}
    \sum_\alpha Q_\alpha V_\alpha^M \approx \sum_\alpha Q_\alpha P_{\alpha 0} Q_0,
\end{equation}
where $Q_0$ is the total charge on the molecule, and
\begin{equation}\label{Palpha}
    P_{\alpha 0} =
    \frac{1}{4\pi\epsilon_0}\int\frac{\ts_\alpha(\vr)\rho_M(\vr')/Q_0}{|\vr-\vr'|}
    da dv',
\end{equation}
which is independent of $Q_0$ in the approximation we use. We shall
write the total molecular energy as
\begin{equation}\label{Emol0}
    E_{tot}(n) = E^{(0)}_{tot}(n) + \sum_\alpha Q_\alpha P_{\alpha
    0} Q_0.
\end{equation}
Here $E^{(0)}_{tot}(n)$ is the total molecular energy, obtained from
the quantum calculations, where the first term in Eq.
(\ref{SchrPotential}) is omitted. In the concrete cases we are
considering in this paper, the gate electrode is reasonably far away
and if the source and drain electrodes are kept at the same or only
small potential differences, then the approximation should be OK.

The rest of the analysis proceeds as in the orthodox theory. The
electrode charges, $Q_\alpha$, are eliminated and we consider two
processes a) and b) that can transfer an electron from source to
drain. The a) process is possible if the electrode potentials
satisfy the inequality
\begin{equation}\label{aprocess}
    eV \geq 2 \frac{\Delta U_a + f_G eV_G}{1+f_S-f_D},
\end{equation}
where
\begin{equation}\label{Ua}
    \Delta U_a =
    E^{(0)}_{tot}(n+1)-E^{(0)}_{tot}(n)-\frac{e^2}{2C'}(2n+1),
\end{equation}
with a different capacitance $C'$ however:
\begin{equation}\label{C'}
    C'^{-1} = \sum_{\alpha\beta} P_{\alpha 0}P_{\beta 0}
    C_{\alpha\beta}.
\end{equation}
The reason for this change in capacitance is, that the large self
interaction of the charge on the island, $P_{00}Q_0 Q_0$, is in the
molecular case contained in the $E^{(0)}_{tot}$ term via the through
the electron-electron interaction term (the last term of Eq.
(\ref{finalCoulombEnergi})). Likewise, the b) process is possible if
\begin{equation}\label{aprocess}
    eV \geq 2 \frac{\Delta U_b - f_G eV_G}{1-f_S+f_D},
\end{equation}
with
\begin{equation}\label{Ub}
    \Delta U_b =
    E^{(0)}_{tot}(n-1)-E^{(0)}_{tot}(n)+\frac{e^2}{2C'}(2n-1).
\end{equation}
Again we obtain a diamond structure like in Fig. 2, and the
fractional charge parameters are determined by the slopes of the
diamond. The width of the charge $n$ diamond is different. It will
be given by
\begin{equation}\label{deltaVg}
    e\Delta V_G = f_G^{-1}
    \left(E^{(0)}_{tot}(n+1)+E^{(0)}_{tot}(n-1)-2E^{(0)}_{tot}(n)\right)-\frac{e^2}{f_G C'}.
\end{equation}
This formula is the important end result of the general analysis. To
summarize the discussion so far, we can say that the orthodox theory
has a very wide domain of applicability, which includes a situation,
where image charge effects are relevant. These image charges are not
to be considered part of the electrodes, but can be viewed as
extensions of the molecule, and be included in the quantum
calculation of the molecular electronic states.

\section{OPV5}
In the past few years there has been a drive to experimentally
realize single molecule transistors. Several strategies has been
used, and in the literature now has reports of many such
transistors\cite{no8}. Here we shall focus on the work of Kubatkin
et al.\cite{no7} who very carefully has grown two gold leads on top
of a gate electrode spatially separated from the rest of the system
by an insulating Al$_2$O$_3$ layer. By in situ monitoring the
source-drain conductance while depositing gold, it is in fact
possible to obtain a source-drain distance of approximately 2 nm.
With this electrode geometry the organic oligo-p-phenylenevinylene
derivative, OPV5, with 5 benzene rings in a chain, which in each end
is terminated by sulfur and a tertiary butyl group. At somewhat
elevated temperatures ($~ 50$ Kelvin) the molecule diffuse on the
surface until a increase of conductance is suddenly observed. This
is interpreted as the realization of an OPV5 forming an electronic
bridge between the source and drain electrodes. The length of OPV5
is 3.2 nm, so this is physically possible. Temperature is lowered to
liquid Helium temperatures, and a full electronic characterization
of the transistor is made. It shows a very beautiful set of no less
than 7 or 8 full SET diamonds. The conductance in the current
carrying modes is well below the conductance quantum, so the theory
of the previous section should apply.

\begin{figure}
\centerline{\includegraphics*[width=70mm]{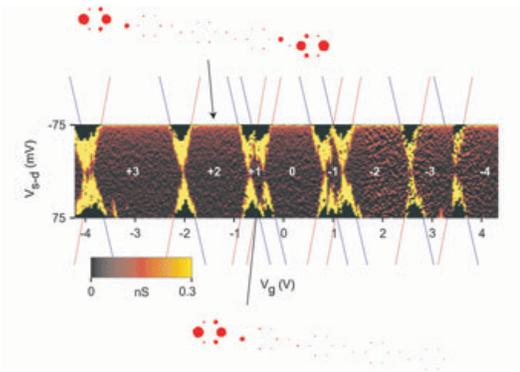}}
\caption{The Coulomb blockade diamonds from Ref. 7. Seen are 9 redox
states of the OPV5 molecule. The two insets in the figure show the
calculated charge densities of the $n=-1$ (charge $Q=+e$) and $n=-2$
states.}
\end{figure}

The first thing to do is to establish the conversion factor, $f_G$,
between the gate voltage axis and energy. It is obtained from the
slopes of the diamonds trough Eq. (\ref{fG}). The fact, that all
observed slopes are identical gives support to the theoretical
picture presented. The analysis gives $f_G = 0.189$.

What is to be expected? First a comparison to electrochemical
experiments is made. These experiments are quite analogous to the
SET experiments. In electrochemistry we have electrons passing from
working to counter electrodes controlled by a third reference
electrode. In solution diffusing OPV5 molecules act as electrolyte
and electron shuttles, carrying electrons between electrodes. The
shuttle only works if a the molecule has an affinity level aligned
with the source electrode Fermi energy. By changing the voltage on
the reference electrode, the electronic chemical potential of the
solution
--- including the OPV5 molecules --- the experiment can map
electronic levels of the molecule. Two sets of redox states are
found, with a quite large gap, 2.5 eV, between them. This is also
the value of the HOMO-LUMO gap measured directly by optical
absorption

As a first attempt at calculating the width of the $n=0$ diamond we
will assume, that the charge on the molecule is uniformly
distributed over the carbon atoms. In this case the width of the
$n=0$ is the HOMO-LUMO gap of the unperturbed molecule. This is a
more than a factor of 10 larger than the observed width, which is
$220$meV. We therefore have to conclude, that severe relaxation
effects are taking place when electrons are added or subtracted the
OPV5 in the present geometry. A possible relaxation mechanism is the
structural deformation of the molecule which takes place when adding
an electron. This, however, only amounts to an energy of
approximately 200 meV \cite{no10}, which has no chance of explaining
the large effect observed.

We are thus led to consider the possibility of direct electrostatic
coupling to the nearby source and drain electrodes. This in turn has
forced us to reconsider the basic electrostatics in a Single
Electron Transistor with a very small central island with quantized
energy levels. This is what was presented in the previous section.
In order to make a quantitative estimate of the size of image charge
effects we need a somewhat realistic model of the molecule, where
these effect can be included without to much effort. We have chosen
to describe OPV5 using the Hubbard model. The important orbitals of
the OPV5 are the 38 $p_z$-orbitals  of the carbon atoms, and the
coupling between nearest neighbor atoms are given by a matrix
element $t$ (or $\beta$ in the quantum chemistry literature). The
value of this is well established, and is $t = 3.9$eV. This
e.g.~reproduce the optical HOMO-LUMO gap. We further need a term
which describe the Coulomb repulsion among electrons on the
molecule. The most important term is here the intra atom repulsion,
which is parameterized by the so-called Hubbard $U$. In the
calculations below, the value is taken to be $U = 4.2$eV. There is
no agreed upon value of $U$ in the literature. It is an effective
parameter, which depends on the chemical environment of the carbon
atom in question.

It is not feasible to do exact calculations, even using the
simplifications of the Hubbard model. We are doing a mean field
calculation, where the spin densities $\sdu$ and $\sdd$ ($i$ being
an atom index) are determined selfconsistently. In the neutral OPV5
molecule far away from all electrodes, the spin densities are
uniform, $\sdu = \sdd = 0.5$, resulting in the single electron
spectrum shown in Fig.~4. We note the HOMO-LUMO gap of 2.5 eV. If an
electron is removed from the molecule, the new densities are still
uniform, and the resulting spectrum is unchanged (except for a
simple energy shift). This is actually a non-trivial result. For
larger $U$'s one finds non-uniform spin and charge densities, which
is a general feature of strongly correlated electron systems.
Calculations for values of $U$ so large, that added (or removed)
electrons form a state with an inhomogeneous charge distribution,
show that the calculated width of the $n=0$ diamond is still several
electron volts (depending on the exact value of $U$). We conclude,
that a molecule unperturbed by the electrodes --- except for a
constant shift of the electrochemical potential --- cannot account
for the observations.

The first attempt of including the effect of the metallic
electrodes, is to model an electrode as a half infinite conducting
plane --- the standard geometry of undergraduate electrostatics. We
have carried out calculations for the a molecule placed with varying
distances and angles with respect to the model electrode. For all
realistic parameters (e.g. distances larger than 1.5 \AA) we find no
inhomogeneous charge distributions, and hence no drastic
renormalization of the $n=0$ diamond width. The reason is, that only
the carbon atom closest to the electrode feel a potential image
charge, and that reduces the effect to negligible changes.

\begin{figure}
\psfigur{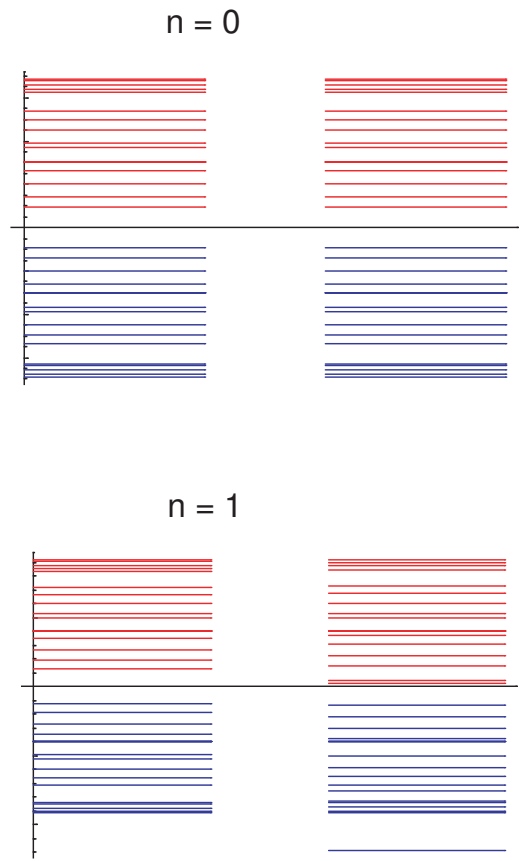}{7} \caption{One-electron spectra for the $n=0$ and
$n=1$ redox states. The left panel is spin up states, the right is
spin down states. Occupied states are blue and unoccupied states are
red.}
\end{figure}

\begin{figure}
\centerline{\includegraphics*[width=70mm]{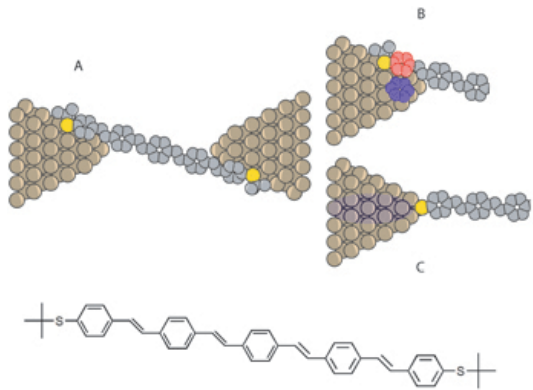}}
\caption{Likely geometries for the OPV5 based Single Molecule
Transistor. See text for explanation. Below is shown the chemical
structure of OPV5.}
\end{figure}

The actual electrodes are in fact heaps of gold atoms, and in the
final geometry we have considered, the benzene rings at the ends of
the molecule is lying flat on the electrode (within approximately a
van der Waals distance), so that a charge imbalance on any of the
six carbon atoms of these rings will couple relatively strongly to
its image charge in the electrodes. Calculations now show a strong
perturbation of the electronic structure of the molecule\cite{no12}.
By varying only the distance between molecule and electrode we are
able to fit the three central diamonds, which involves $n=-2, -1, 0,
1, 2$ redox states of the molecule. The actual distance between the
benzene ring and the mirror plane of the gold electrodes chosen to
fit the data is 2.2 \AA, which is quite realistic, given the known
van der Waals radius of benzene. In figure 3 we also show the
electron densities of some of these states. For the $n=1$ state the
charge pile up at the end benzene ring of the molecule. The $n=2$
state, which is quite close in energy, the two charges occupy each
end of the molecule. There are two such states, but it turns out
that the singlet state (spin = 0) is approximately 70 meV lower in
energy, so we propose that this is the relevant state to the
experiment. The one electron energy spectra, which enters the
calculations, are shown in Fig.~4. We see, that they are completely
scrambled, and no intuition is to be gained by only considering
these spectra. Only the total energies carry the relevant
information.

\section{Conclusion}
In conclusion we find, that the image charge effects can be involved
in the physics of Single Molecule Transistors. It requires, however,
a special geometry. By varying the end groups of the molecule (and
other molecules of the OPV family) we expect a geometry change that
will remove the strong perturbation due to image charges. If e.g.
the tertiary butyl group is removed so that the Sulfur atom binds
directly to the Gold electrodes the possibility of having the end
benzene ring couple electrostatically to the electrode is severely
impeded and the $n=0$ diamond should become approximately a factor
of 10 greater corresponding to the unrenormalized HOMO-LUMO gap.
These different situations are summarized schematically in Fig.~5,
where Fig.~5A - B show a likely orientation of the molecule in the
device realized in the experiment by Kubatkin et al.~in which the
terminal thiol capped benzene ring is protected by a tertiary butyl
group. Fig.~5B illustrate the image charge as a blue shadow of the
``red'' charges in the molecule. In this case all of the atoms in
the terminal benzene ring are affected by image charges and
calculations show a strong localization of the charge on this ring,
as also seen in Fig.~3. The van der Waal's nature of the contact
between the physisorbed molecule and the electrode, with the
hydrogen atoms at the edge of the benzene ring being the likely
contact region, prevent a strong electronic coupling between the
electron system and the electrode. In turn this results in a
relatively large tunneling barrier and Coulomb blockade behavior as
experimentally observed\cite{no7}. Shown as Fig.~5C is the likely
orientation of the same terminal benzene when it is deprotected
allowing the terminal thiol to form a chemical bond to the gold
electrode. In this case the benzene ring is pushed away from the
surface of the electrode by the covalent thiol bond. Our
calculations show that image charge effects are seriously impeded in
this case, and we therefore predict that these effects will play a
minor role. Interestingly, the electronic coupling between the
electrode and the molecule, now mediated by a covalent bond, is
expected to be stronger then the coupling mediated by the van der
Waals contact discussed above.

Taken together the above analysis reveal an intricate relation
between the details of the molecule-electrode contact and the
resulting effective charge transfer channels. The relation resembles
the competition between charge delocalization and localization often
discussed in terms of the Hubbard parameters $U$ and $t$. In our
case, however, the electron localization is driven by image charge
effects rather than electron-electron repulsion. In the special case
described in the present paper, it is shown that image charges can
influence this electronic instability in direction of the localized
regime very significantly. The key parameter controlling the
magnitude of this effect is the distance of the atoms in the
molecule to the electrode surface. This parameter is independent of
the degree of electronic overlap between the electrons in the
molecule and electrode, which tends to favor coherent transport
through delocalized states. A series of experiments in which these
two parameters are varied independently are currently under
investigation in our laboratories, and we expect a series of new
phenomena to be reveled as we systematically map the properties of
single molecule devices in which both delocalized coherent transport
and sequential electron hopping between localized states are active.


\begin{thebibliography}{10}
\bibitem{no12} 
In the calculations presented in this paper, we have not made a
complete calculation of the kernel $K(\vr,\vr')$. Rather we have
made a constant shift of the atomic energies of all the orbitals of
the first (or last) bezene ring. The shift is proportional to the
net charge on the benzene ring, and the distance dependence is
determined by a simple model, where a ring of charge is placed above
a metallic plane. Such an approximation will have all the correct
features of the exact kernel $K(\vr,\vr')$.
\end{thebibliography}
\end{document}